# Universal pulses for homogeneous excitation using single channel coils

Ronald Mooiweer[1,2,3], Ian A. Clark[4], Eleanor A. Maguire[4], Martina F. Callaghan[4], Joseph V. Hajnal[1,2], Shaihan J. Malik[1,2]

[1]*School of Biomedical Engineering and Imaging Sciences, Faculty of Life Sciences and Medicine, King's College London, United Kingdom*

[2]*Center for the Developing Brain, School of Biomedical Engineering and Imaging Sciences, King's College London, St. Thomas' Hospital, London, United Kingdom*

[3]*MR Research Collaborations, Siemens Healthcare Limited, Frimley, United Kingdom*

[4]*Wellcome Centre for Human Neuroimaging, UCL Queen Square Institute of Neurology, University College London, London, United Kingdom*

Corresponding author:

Dr. Shaihan Malik

Email: shaihan.malik@kcl.ac.uk

## Abstract

**Purpose**

Universal Pulses (UPs) are excitation pulses that reduce the flip angle inhomogeneity in high field MRI systems without subject-specific optimization, originally developed for parallel transmit (PTX) systems at 7T. We investigated the potential benefits of UPs for single channel (SC) transmit systems at 3T, which are widely used for clinical and research imaging, and for which flip angle inhomogeneity can still be problematic.

**Methods**

SC-UPs were designed using a spiral nonselective k-space trajectory for brain imaging at 3T using transmit field maps ($B_1^+$) and off-resonance maps ($B_0$) acquired on two different scanner types: a 'standard' single channel transmit system and a system with a PTX body coil.

The effect of training group size was investigated using data (200 subjects) from the standard system. The PTX system was used to compare SC-UPs to PTX-UPs (15 subjects). In two additional subjects, prospective imaging using SC-UP was studied.

**Results**

Average flip angle error fell from 9.5±0.5% for 'default' excitation to 3.0±0.6% using SC-UPs trained over 50 subjects. Performance of the UPs was found to steadily improve as training group size increased, but stabilized after ~15 subjects.

On the PTX-enabled system, SC-UPs again outperformed default excitation in simulations (4.8±0.6% error versus 10.6±0.8% respectively) though greater homogenization could be achieved with PTX-UPs (3.9±0.6%) and personalized pulses (SC-PP 3.6±1.0%, PTX-PP 2.9±0.6%).

MP-RAGE imaging using SC-UP resulted in greater separation between grey and white matter signal intensities than default excitation.

**Conclusions**

SC-UPs can improve excitation homogeneity in standard 3T systems without further calibration and could be used instead of a default excitation pulse for nonselective neuroimaging at 3T.

## 1 Introduction

Inhomogeneity of the radio frequency transmit field ($B_1^+$) reduces the quality of images obtained at main magnetic field strengths of 3T and higher (1,2). $B_1^+$-mitigating techniques have predominantly been developed for Ultra High Field (7T and higher), involving parallel transmission (PTX), radio frequency pulse design and combinations of both (3). These techniques are often based on subject-specific, measured, field distributions and thus require some time and effort to tailor the solution to the subject in the scanner. Recently however, the universal pulse (UP) approach has been introduced (4), which exploits similarity in field distributions across subjects. UPs were calculated based on the field maps of a training group of subjects, after which they could be applied to other subjects to retrieve the excitation homogeneity. In general, UPs yield better homogeneity than standard RF pulses but do not achieve the same performance as fully subject-specific pulses. This drawback is offset by avoiding spending time on acquiring field maps and doing calculations for every individual subject, making UPs suitable for standard clinical workflow.

UPs were originally considered for PTX coils at 7T (4) and have been extended from T1 weighted imaging to 3D Turbo Spin Echo sequences (5) and gradient echo EPI (6). 'Universal' (i.e. group optimized) solutions have also been shown to improve performance for the Direct Signal Control (dynamic RF shimming) PTX method applied to 3D FLAIR imaging at 7T (7) and for body imaging at 3T (8). While not as severe as at 7T, $B_1^+$ inhomogeneity is also still an issue for 3T neuroimaging both for quantitative methods and structural imaging requiring automated segmentation. The objective of the work reported here is to introduce calibration-free universal pulses for homogeneous excitation within T1 weighted sequences for neuroimaging at 3T using standard clinical systems that use single channel (SC) excitation. Important questions to be addressed before UPs can be applied widely with confidence are the minimal training group size needed, and the performance on a large range of subjects. Previous studies with UPs showed marked improvement compared to default excitation, often using relatively small (6-20) sample sizes for training (4,9–11).

In this study, universal pulses were designed and evaluated on a large set of field maps that were acquired as part of a larger neuroscience study (12,13) using standard 3T scanners, where $B_1^+$ maps were used to correct quantitative MRI maps. Although most previous work on UPs used adaptations of k-t point RF pulse design (14), this work used the spiral nonselective excitation (SPINS) trajectory which has already been shown to reduce excitation inhomogeneity at 3T both using PTX and SC excitation (15). A prototype PTX-capable 3T scanner was also used to investigate differences in performance between SC-UPs and PTX-UPs, and to perform a prospective imaging demonstration on two subjects. Aspects of this study were presented at the annual meeting of the ISMRM in Honolulu, HI, in 2017 (16).

## 2 Methods

### 2.1 UP design using SPINS

SPINS pulse design uses the spatial domain method (17) which may be written as a matrix problem

$$\mathbf{m} = \mathbf{A}\,\mathbf{b}, \qquad [1]$$

where **m** is the target magnetization in the subject, **A** is a system matrix constructed based on the subject's $B_1^+$ and $B_0$ field maps, the spatial coordinates of the voxels, and the prescribed 3D non-selective k-space trajectory. Lastly, **b** is the complex RF pulse to be determined.

Throughout this work we used the k-space trajectory proposed in ref (15), after correcting for measured imperfections in the impulse response function of the gradient system (18,19), with an overall duration 1.37 ms (0.37 ms gradient-only lead-in and 1 ms with jointly activated RF). To find the RF pulses that excite a spatially uniform target magnetization **m**, **b** is optimized using a magnitude least-squares approach (20):

$$\boldsymbol{b} = argmin_b \{\||\boldsymbol{Ab}| - \boldsymbol{m}\|^2 + \lambda\|\boldsymbol{b}\|^2\}, \qquad [2]$$

where λ is a Tikhonov regularization factor.

To construct a *Universal* SPINS pulse from a training group of N subjects, a combined matrix description of the problem is constructed by concatenating the individual matrices:

$$\begin{bmatrix} \boldsymbol{m}_1 \\ \vdots \\ \boldsymbol{m}_N \end{bmatrix} = \begin{bmatrix} \boldsymbol{A}_1 \\ \vdots \\ \boldsymbol{A}_N \end{bmatrix} \boldsymbol{b}_U, \qquad [3]$$

$$\boldsymbol{m}_U = \boldsymbol{A}_U \boldsymbol{b}_U.$$

This is then solved using the same magnitude least-squares algorithm to yield a single universal RF pulse waveform **b**$_U$. The Tikhonov regularization factor was set to scale with the number of transmit channels and number of subjects in the training group, to reflect the increasing dimensions of the design problem.

### *2.2 Design considerations of UPs: simulation study*

Two-hundred healthy volunteers were included in the study. They were aged between 20 and 41 years old (mean age 28.7 years, standard deviation = 5.6 years), and reported no psychological, psychiatric, neurological or behavioural health conditions. The subjects were scanned over two years and included 98 males and 102 females. The $B_1^+$ and $B_0$ maps were collected as part of larger study (12,13) for which subjects were reimbursed £10 per hour for taking part. All subjects gave written informed consent and the study was approved by the University College London Research Ethics Committee.

These data were collected using three identical 3T scanners using SC RF transmission (Tim Trio, Siemens Healthcare, Erlangen, Germany), so the data could be pooled without distinguishing between the scanners. In all subjects, the prescribed field of view was the same and no instructions were given to the subjects regarding tilting of the head in the sagittal plane.

$B_1^+$ maps were acquired using a stimulated echo/spin echo sequence with a 3D EPI read-out (21). EPI-related distortions were corrected using acquired $B_0$ maps (2,22). The FSL-BET brain extraction tool (23) was used to define a region of interest (i.e. the whole brain) in the field maps that was then used for pulse design. For pulse design and subsequent simulations, the maps were down-sampled to a grid of 32 x 32 x 32 voxels (8 x 6 x 6 mm$^3$).

Universal SPINS pulses were calculated as described in the previous section, using the $B_1^+$ and $B_0$ maps of a number of subjects in a training group, and evaluated on a test group using Bloch Equation Simulations (24). One-hundred subjects were randomly chosen as the test group on which UPs with various design parameters were evaluated. This test group was fixed across all simulations to facilitate a consistent comparison between design methods. The effect of the training group size ($N_{training}$) on the efficacy of UPs was evaluated using simulation. In each case, UPs were designed using $B_1^+$ and $B_0$ maps from $N_{training}$ subjects randomly selected from the 100 not included in the test group. This process was repeated 20 times for each value of $N_{training}$ to avoid selection bias and to study the variability/stability with respect to the random selection of training subjects, always evaluated on the same testing group.

A possible improvement to the design of universal SPINS pulses was considered, addressing natural variation in the positioning of a subject's head in the scanner. The transmit field is expected to vary in a relatively predictable way in relation to changes in the position of the head with respect to the coil. We therefore attempted to make the SC-UP more effective by applying a phase correction term $\phi(t) = \boldsymbol{k}(t).\Delta\boldsymbol{r}$ to the RF pulse based on an individual's head position within the coil, with $\Delta\boldsymbol{r}$ the vector of the brain center of mass from the origin of the coordinate system, determined from a mask of the brain. When using this method, UPs were calculated after first centering the coordinate system for each training dataset (i.e. by setting $\Delta r = 0$ for each subject). This is possible because in the concatenated UP design approach (Eq. [3]) each subject has their own system matrix **A** which contains the coordinate system within it.

## 2.3 Comparison of SC and PTX UPs: simulation study

A different 3T scanner (Achieva, Philips Medical Systems, Best, The Netherlands) with an eight-channel PTX body coil (25) was used to compare SC and PTX performance in UPs. The quadrature mode of this system ('single-channel mode') is similar to that of a whole body birdcage coil (25). By either designing pulses for the individual channels or the combination of channels driven in quadrature mode, a direct comparison of UPs for PTX and SC body coils could be made. One of the transmit channels was excluded from all experiments (including SC mode) due to a technical failure, resulting in a system capable of 7-channel PTX. An eight-channel head coil was used throughout for signal reception.

For the comparative simulation study, field maps were acquired in 15 healthy volunteers (9 male, 6 female). They were aged between 23 and 35 years old (mean age 28.8 years, standard deviation = 3.7 years). All subjects gave written informed consent and the study was approved by London Riverside Research Ethics Committee. Multi-channel $B_1^+$ maps were acquired using an interferometric approach (26,27) with a combination of low flip angle spoiled gradient echo (SPGR) (nominal flip angle 1°, TR = 3.5 ms, TE = 1.5 ms) for each linear combination of channels, and a single 3D actual flip angle imaging (AFI) (28) map (nominal flip angle 80°, TR = 30, 150ms, TE = 4.6 ms) in quadrature mode. The SPGR images were used to compute relative $B_1^+$ and the AFI map scales to the correct absolute level – a similar approach was used in refs (29,30). $B_0$ maps were acquired using dual echo SPGR (ΔTE = 2.3 ms). Field maps were acquired at isotropic resolution 3.91 x 3.91 x 3.91 mm$^3$ and down sampled to a matrix of size 32 x 32 x 32 (8 x 8 x 8 mm$^3$). For each subject, the FSL-BET brain extraction tool (23) was used to define the whole brain as region of interest for pulse design.

Universal pulse simulations were carried out using a leave-one-out method: UPs were calculated using the field maps of 14 subjects and evaluated on the remaining subject, this was then repeated for every subject. Personalised pulses (PP) were also designed for SC and PTX operation of the body coil using the conventional SPINS method.

## 2.4 Data analysis: pre-normalized NRMSE

To compare the homogeneity of the different excitation strategies, the normalized root mean square error (NRMSE) of the simulated excitation profile was calculated over the brain. Before determining the NRMSE, the flip angle distributions were normalized such that their mean value always matched the desired angle. Although an overall offset in flip angle might occur, this could be corrected in practice by appropriate amplitude scaling, as is commonly done for conventional RF excitation. By applying this pre-normalization, we ensured that the NRMSE only reflected the variation in flip angle per subject and that NRMSE's across different subjects could be compared.

## 2.5 MP-RAGE imaging with SC-UP: imaging study

The same PTX-enabled system was also used to demonstrate SC-UPs in MP-RAGE imaging of two additional healthy volunteers (1 male, 1 female, aged 25 and 27 years old) who gave written informed consent for this study that was approved by the London Riverside Research Ethics Committee. The Alzheimer's Disease Neuroimaging Initiative MP-RAGE protocol (31) was used with both standard RF pulses, and after replacing the small flip angle excitation pulses (flip angle 8°) with UPs. The SC-UP was calculated from field maps of the 15 subjects scanned on this system. As previous work has shown that tailored RF pulses at flip angle 8° can improve excitation homogeneity but not contrast (15), MP-RAGE imaging was also performed with SC-UP at a flip angle of 5.5°, the effective flip angle that is achieved in the cortex with quadrature mode excitation. The images were first brain extracted and then evaluated by segmentation into three tissue classes (nominally white matter, grey matter and CSF) with simultaneous bias field correction using FSL FAST (32).

## 3 Results

### *3.1 Design considerations of UPs: simulation study*

The results of varying the training group size are summarized in **Figure 1A**, where every entry in the graph is based on the simulated performance on 100 test subjects, repeated 20 times for different selections of the training group. Remarkably, even for a training group size of 1 the mean NRMSE of the SC-UP (4.0±1.2%) is far below that of default excitation (9.5±0.5%). Increasing the group size brings down the average NRMSE: 3.1±0.6% for $N_{training}$=15 and 3.0±0.6% for $N_{training}$=50. When any 15 or more subjects are chosen, the worst NRMSE over all test subjects is below 6.5%, and 80% of the simulations are within ~1% NRMSE of the average value. **Figure 1B** shows the standard deviation of the NRMSE across the 100 different testing subjects, averaged over the 20 different instances of training group. It shows that a spread of ~0.6% NRMSE is present for any training group size (5-50) in these experiments, due to the differences in properties of the 100 testing subjects. In contrast, in **Figure 1C,** the average standard deviation across repetitions of the training instances decreased with an increasing number of training subjects. Low values here (~0.2% NRMSE for 15-20 subjects) show that the performance of the UP is stable no matter which training subjects are chosen as long as there are a sufficient number. The best results were obtained with 50 subjects inside the training group, but the improvement is small compared to $N_{training}$=15.

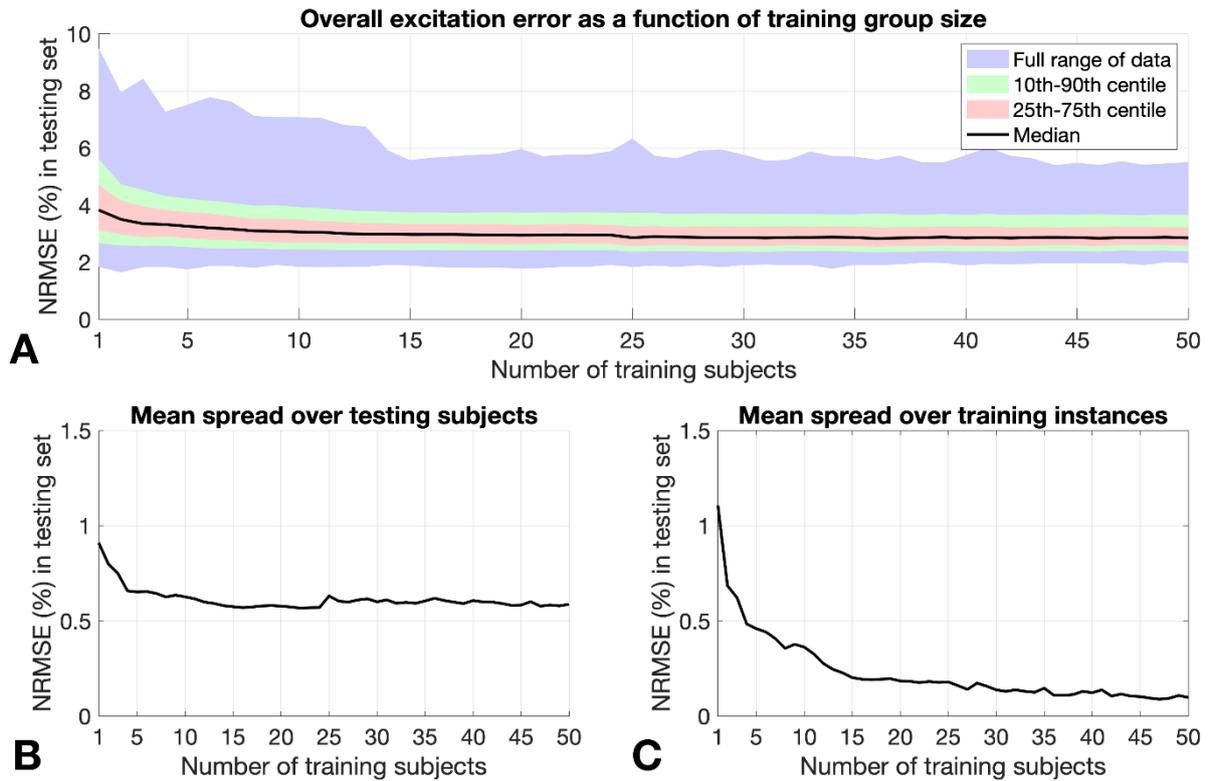

*Figure 1. **A)** Simulated excitation error as NRMSE of SC-UPs is optimised using 1-50 training subjects, evaluated on 100 testing subjects, and repeated 20 times with different selections of training subjects. Although the total spread per experiment with a fixed training group size is 4% NRMSE or larger, the majority of the simulations are within a much narrower range of ~1% NRMSE. **B)** Mean, over training group instantiations, of the standard deviation of the SC-UP excitation NRMSE evaluated across the 100 testing subjects. Aside from the very smallest training group size, the standard deviation is generally constant, reflecting the variation in test subjects' field maps and the resulting excitation errors over the test subjects. These remain around 0.6% NRMSE no matter how much data is included in the training group size. **C)** Mean, across all test subjects, of the standard deviation of the SC-UP excitation NRMSE across the 20 independent training instances of the same training group size. The variable resulting NRMSE as a function of selecting different training subjects drops rapidly until ~15 after which it reduces very gradually with increasing training group size. This suggests that once a training group size of 15 is reached, the results of SC-UPs are largely independent of which specific subjects are used for training.*

Possible improvements of brain centering are shown for a 'stable' training group size of 15 in **Figure 2**. This demonstrates that, even though a modest average improvement was obtained by using brain centering in UPs, not all subjects benefitted from this method.

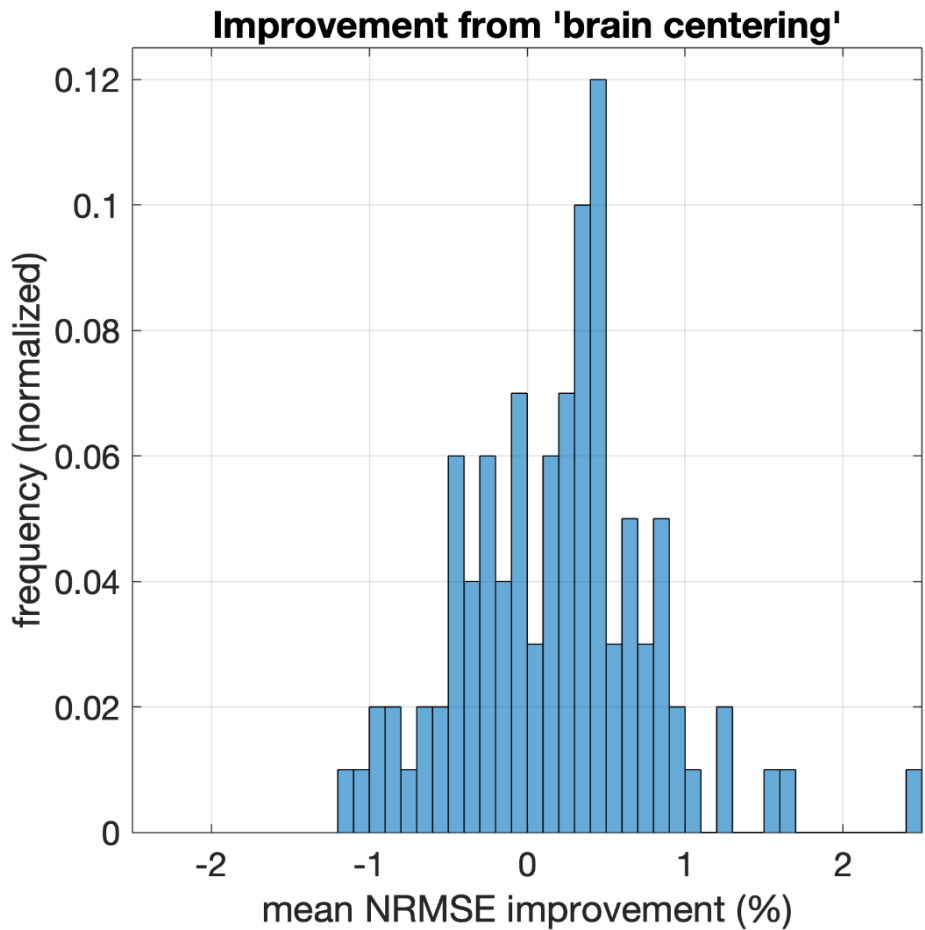

*Figure 2.* *Improvement due to brain centering over all 100 test subjects. Brain centering is mostly, but not always, an improvement. The average improvement across subjects is 0.19 percent.*

In **Figure 3** the performance of SC-UPs ($N_{training}$ = 15, brain centering enabled) was compared to default excitation and personalized single channel SPINS pulses (SC-PP). Tailored pulses performed best, with a small spread, but SC-UPs also provided a large improvement versus the default excitation, and performed better than default excitation in all cases.

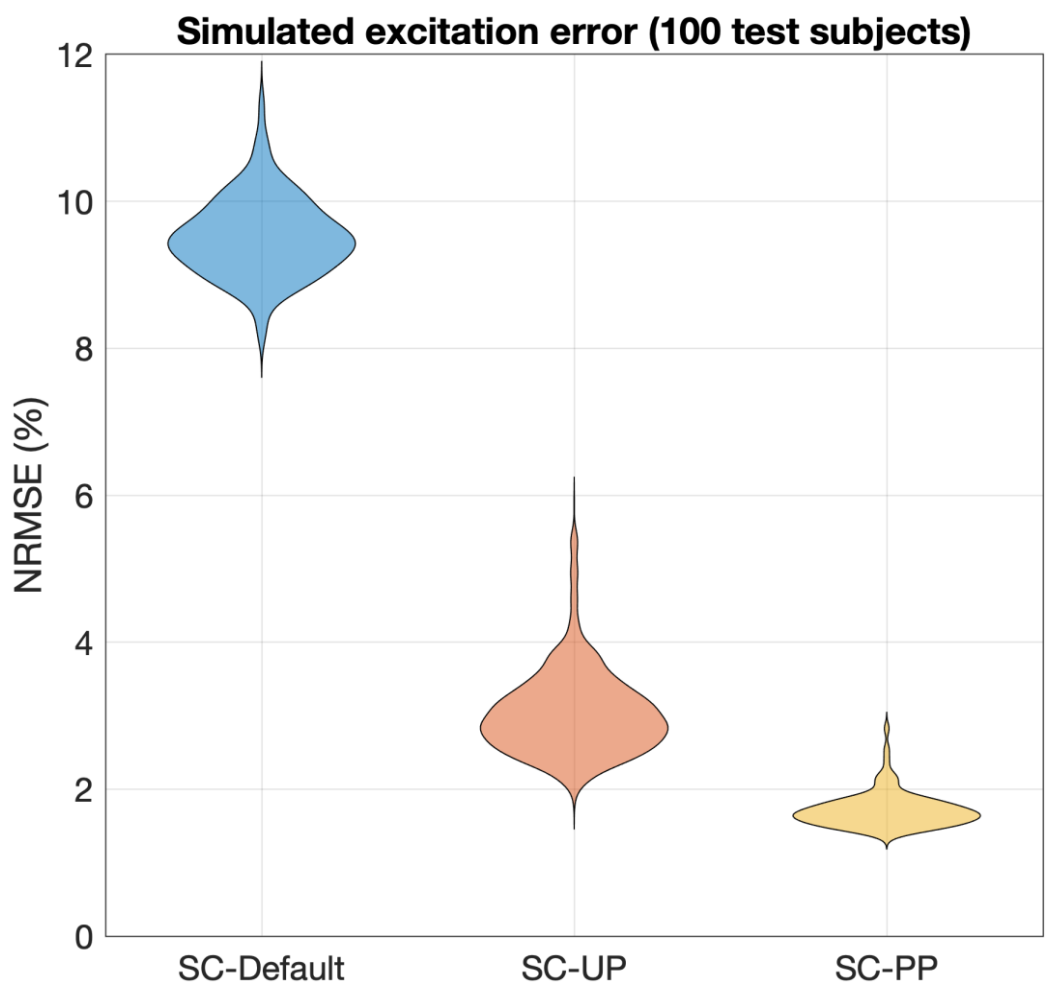

*Figure 3. Distributions of simulated NRMSE values for single channel excitation using the default non-selective pulse, SC-UP (Ntraining = 15, brain centering enabled, repeated 20 times with new selection of training subjects), and SC-PP (personalised SPINS pulse). There is no overlap between the distributions of default excitation and SC-UP, indicating that SC-UPs always outperformed defualt excitation in terms of NRMSE. Mean and standard deviations are: default 9.5+/-0.5, SC-UP 3.0+/-0.6, SC-PP 1.7+/-0.2.*

## 3.2 Comparison of SC and PTX UPs: simulation study

In the simulations for the PTX-enabled system, the SC-UP again performed consistently better than the default quadrature mode (**Figure 4**), reducing the mean NRMSE from 10.6±0.8% to 4.8±0.6%. Personalized designs (SC-PP, PTX-PP) and using PTX for UP design tended to reduce the excitation error somewhat further (3.6±1.0%, 2.9±0.6% and 3.9±0.6%, respectively), as expected.

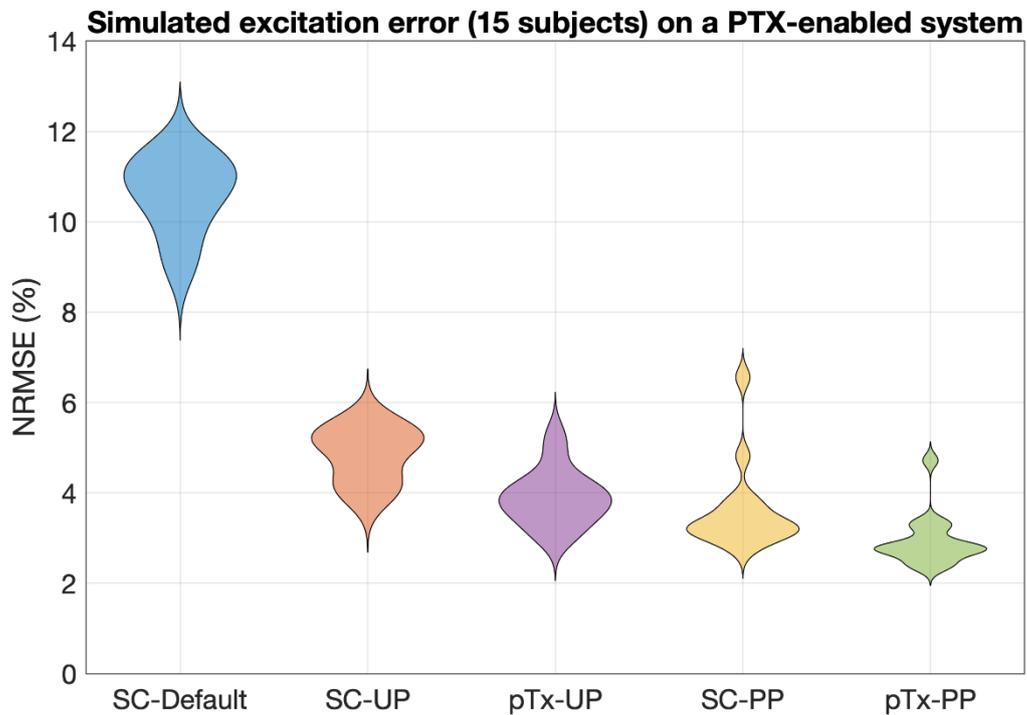

*Figure 4.* Excitation variations simulated for the PTX-enabled 3T scanner in 15 subjects using default, SC-UP, PTX-UP, SC-PP, and PTX-PP. SC-UPs always resulted in lower excitation error than default excitation. Additional reductions were achieved by using PTX and PP.

## 3.3 MP-RAGE imaging with SC-UP: imaging study

The SC-UP, calculated using the field maps of all 15 previously-acquired datasets on the PTX system, is shown in **Figure 5**. Post-bias field correction MP-RAGE images for 1 volunteer are shown in **Figure 6 (top)**. The contrast in the image acquired with SC-UP at 8° flip angle didn't improve compared to default excitation: although a greater excitation homogeneity can be expected by the use of UP, the 8° flip angle is known to generate insufficient contrast in MP-RAGE images. However, the MP-RAGE image acquired with SC-UP at 5.5° flip angle does display better contrast, also demonstrated by a larger separation between the white matter and grey matter peaks in the histograms of both volunteers (**Figure 7**). In the data acquired with 5.5° flip angle, a better segmentation of the deep grey matter tissue was achieved using the UP compared to default (**Figure 6, bottom**), also suggesting that the signals obtained with SC-UP are more in line with the ideal tissue contrast that is expected in the Gaussian mixture model used by FSL-FAST.

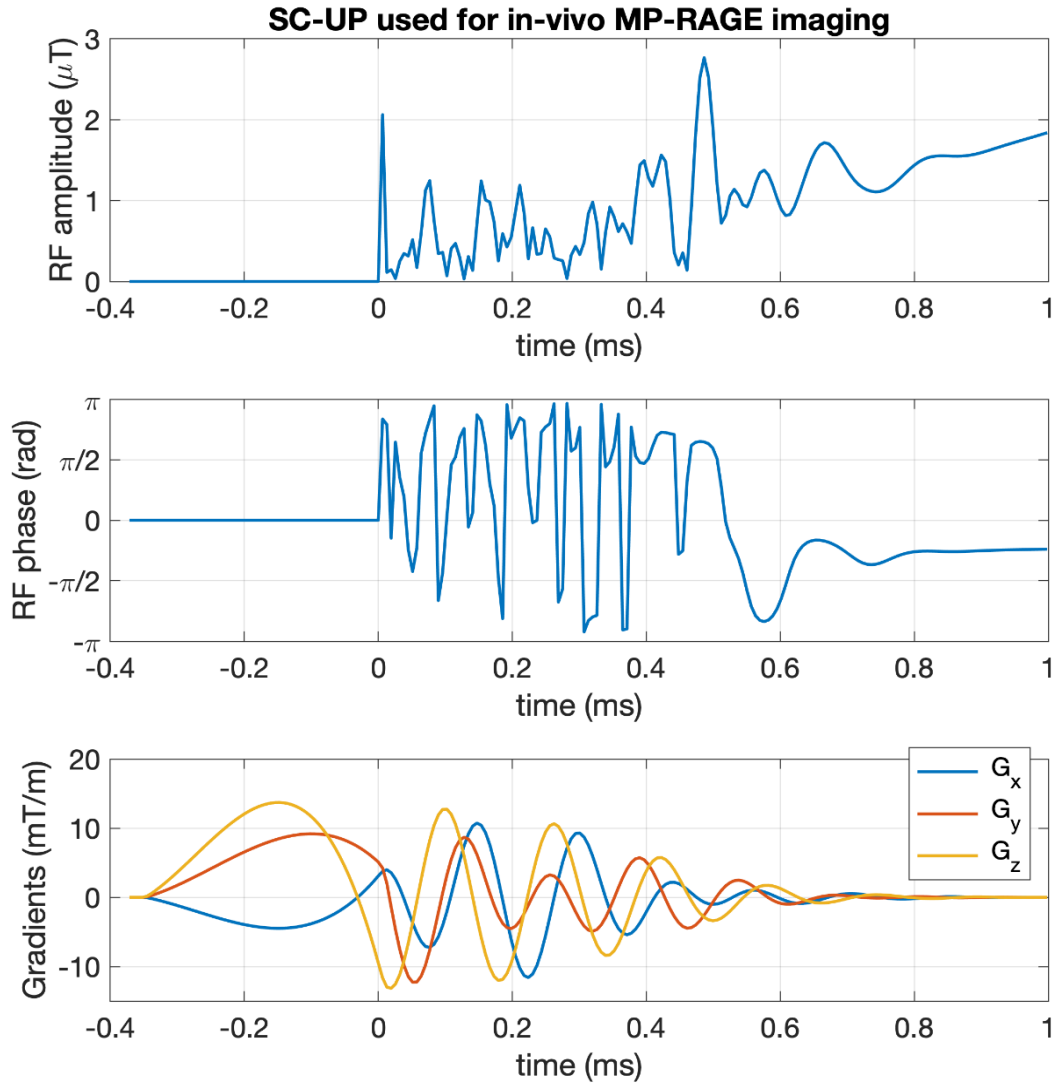

*Figure 5. RF and gradient waveforms of the single channel universal SPINS pulse (SC-UP), optimized using 15 training subjects, which was used for MP-RAGE imaging in 2 different healthy volunteers. RF amplitude was scaled to provide a 5.5° excitation. The gradient waveforms shown are the desired waveforms, before pre-emphasis for the system response using the measured GIRF.*

## 4 Discussion

This paper investigated the use of a universal pulse for homogeneity correction of 3T brain imaging with body coil excitation. Simulations using a large set of field map data acquired from 200 subjects (100 used for testing) showed that SC-UP outperformed standard quadrature excitation, achieving an average NRMSE of 3.1±0.3% with 15 training subjects, compared to 9.5±0.5% with default SC excitation in simulation. A training group size of 15 is recommended as this number marked a drop in the spread in the variability of the simulated results. For both individuals in the acquired imaging experiments and in all simulations (with a large enough training group size) the excitation performance for the UP was never worsened in the test subjects.

When used prospectively for MP-RAGE imaging, SC-UPs yielded good image quality and clearly showed improved tissue contrast. As in the original publication on SPINS (15), it was found that improved contrast for the cortex could only be achieved after reducing the target flip angle. This can be explained by the fact that when using 'default' excitation pulses, the flip angle that is actually achieved in the cortex is much lower than the specified value. Consequently, if this excitation is made uniform, a higher flip angle (with lower contrast) is actually achieved there. Similar behaviour was observed in the present study, with **Figure 7** and the segmentations on **Figure 6B** confirming that the 5.5° SC-UP yielded the greatest difference between grey and white matter peaks and best performing tissue segmentation.

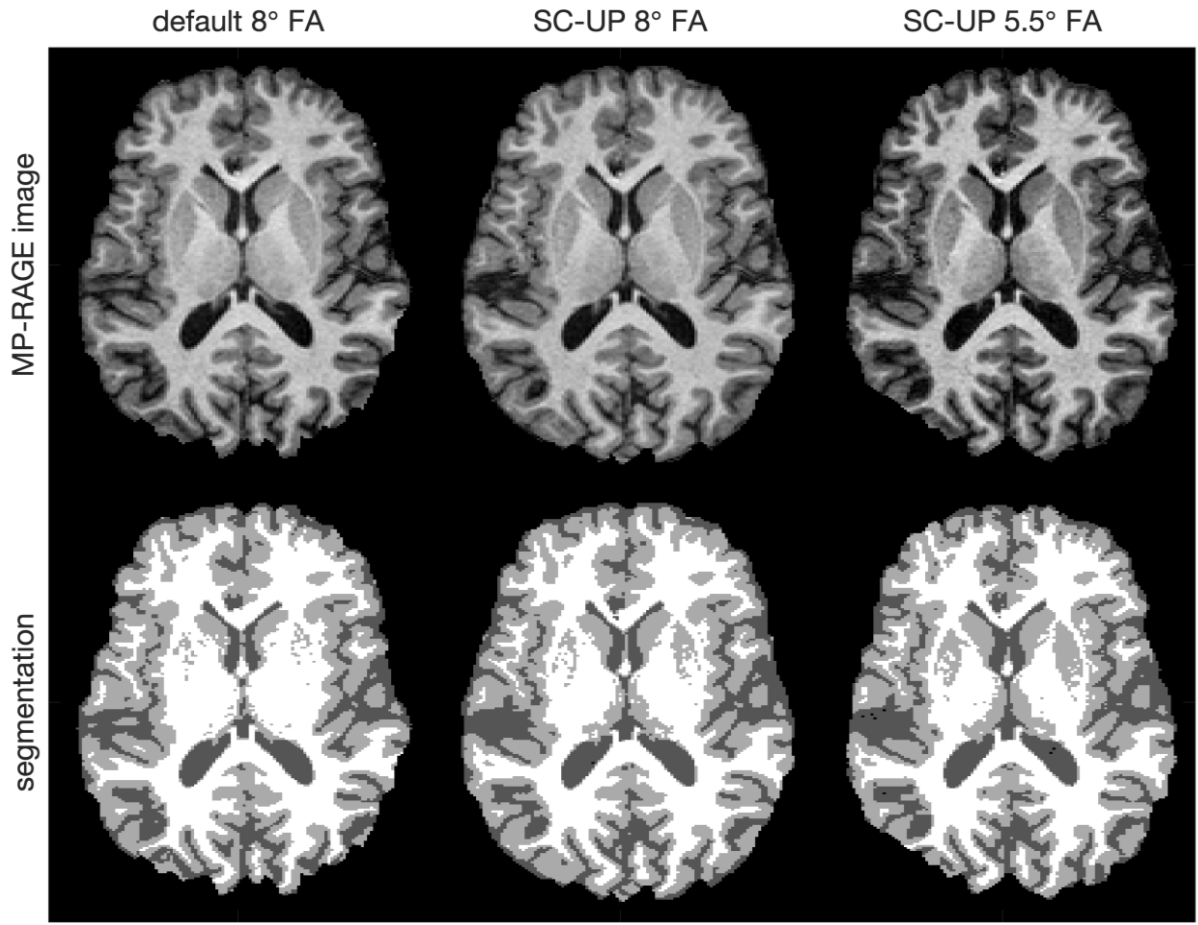

*Figure 6. Top row: Bias field corrected MP-RAGE images acquired in one subject using default excitation and SC-UPs at FA 8° and 5.5°. Bottom row: corresponding three tissue segmentation results after bias field correction using FSL FAST. Superior contrast resulting in better delineation of deep matter structures such as the putamen can be observed when using SC-UP at 5.5° FA.*

In the simulation experiments presented here, the NRMSE on average was reduced from 9.5±0.5% to 3.0±0.6% with UPs while personalised pulses could reach 1.7±0.2%. These results replicate the trend of the original UP work, which reported errors of ~28, ~11 and ~7 % respectively (4). The difference in absolute values may arise from one, or a combination of, the many differences such as field strength, coil type, RF pulse design method, and inclusion of the training group in the simulation results.

Simulation results from the PTX system were similar to those from the more standard SC 3T systems, although the error in the 'default' (i.e. quadrature) mode was higher, likely because the PTX system was used with only 7 of the 8 transmit channels due to hardware failure. Nevertheless, we observed the expected hierarchy of performance, with PTX in general outperforming single channel operation, and personalized pulses generally outperforming universal ones. That said, switching from SC-Default to SC-UP yielded the largest drop in NMRSE for both the regular SC scanner and the PTX scanner.

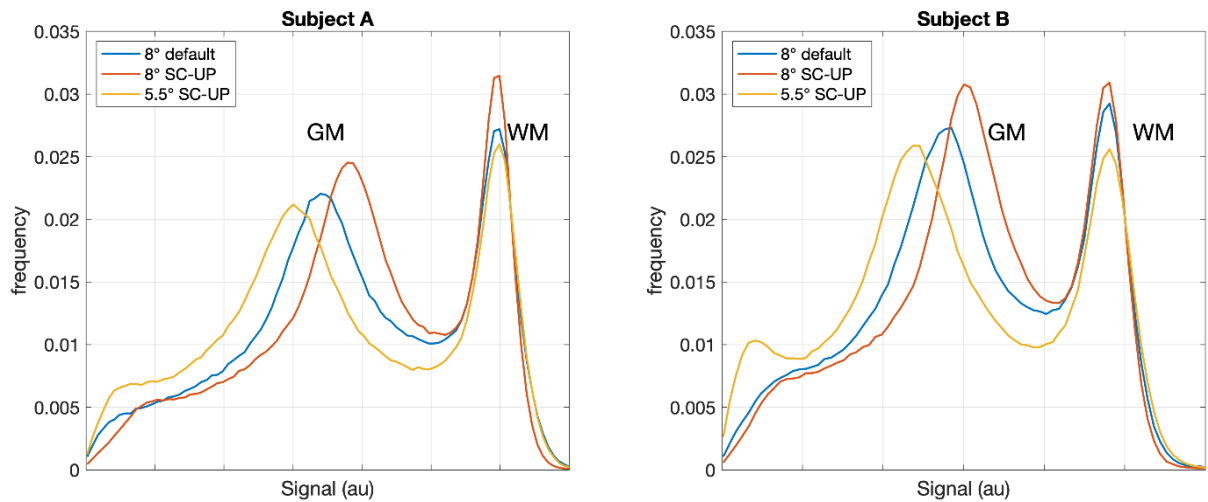

*Figure 7.* Signal intensity histograms of MP-RAGE images acquired in two subjects after bias field correction, using default excitation and the SC-UP at flip angles of 8° and 5.5°. Compared to default excitation, using SC-UP at 8° reduced the gap between grey (GM) and white matter (WM). However, when the SC-UP was used at the optimised 5.5° flip angle, the separation between the GM and WM improved.

The original SPINS method was adapted only minimally to generate UPs for MP-RAGE imaging at 3T, the application for which the SPINS pulses were originally designed. A similar approach could be attempted for other imaging applications, but in such a case alternative pulse design methods could be investigated. For example, the specific absorption rate (SAR) was not explicitly taken into account in the current work since the small flip angle pulses being designed have very low energy. Such a consideration could be added in the future, particularly if high flip angle pulses are desired. Likewise a gradient impulse response function (GIRF) was used to map from applied gradients to k-space trajectory for pulse design. Since the k-space trajectory was held fixed throughout the study this could in practice be calibrated in advance either by using a GIRF, field probes, or other direct measurement. We have observed performance to remain stable over long periods of time (15).

An additional hypothesis tested in this work was that the performance of UPs might be improved by accounting for differences in the positioning of the head relative to the scanner. This could be realized in practice by determining the position of the center of the head from a localizer scan, for example. In order for this correction to be effective, the $B_1^+$ distribution would need to 'move with the subject' – i.e. the center brightening effect observed at 3T would need to remain fixed relative to the position of the subject's head as it moves within the RF coil. This is likely to be a better assumption at 3T using a body transmit coil than at Ultra High Field using a localized transmitter. Even so, while a small benefit was observed from correcting the pulses for head position, it did not improve the results for ~40% of the subjects.

## 5 Conclusions

Single Channel Universal Pulses using SPINS have been shown to be suitable for use in low flip angle (GRE) sequences using non-selective excitation to improve excitation homogeneity without spending time on subject-specific pulse design. Given that there is no intrinsic cost to using these methods, and that they may improve data quality and improve the performance of image post-processing methods, SC-UPs could be used on standard 3T MR systems in clinical and investigational neuroimaging.

## 6 Acknowledgements

This work was supported by the EPSRC (EP/L00531X/1), Medical Research Council (MR/K006355/1), a Wellcome Principal Research Fellowship to E.A.M. (01759/Z/13/Z), a Wellcome Strategic Award to the Wellcome Centre for Human Neuroimaging (203147/Z/16/Z), Wellcome/EPSRC Centre for Medical Engineering (WT 203148/Z/16/Z), and the National Institute for Health Research (NIHR) Biomedical Research Centre at Guy's and St. Thomas' NHS Foundation Trust and King's College London. We thank Anna Monk, Victoria Hotchin and Gloria Pizzamiglio for assistance with data collection. The views expressed are those of the authors and not necessarily those of the NHS, the NIHR or the Department of Health.

This research was funded in whole, or in part, by the Wellcome Trust [Grant numbers: 01759/Z/13/Z; 210567/Z/18/Z; 203147/Z/16/Z; 203148/Z/16/Z]. For the purpose of Open Access, the authors have applied a CC BY public copyright licence to any Author Accepted Manuscript version arising from this submission.

Data from the 200 volunteers are part of a larger project, the data from which will be made available once the construction of a dedicated data-sharing portal has been finalised. In the meantime, requests for the data can be sent to e.maguire@ucl.ac.uk.